\documentclass{webofc}
\usepackage[varg]{txfonts}   % Web of Conferences font
\woctitle{13th International Conference on Meson-Nucleon Physics and the 
Structure of the Nucleon}
\begin{document}
\title{A Vision of Hadronic Physics}
%
% subtitle is optional
%
%%%\subtitle{MENU2013}

\author{Anthony W. Thomas\inst{1} %\fnsep
%\thanks{\email{Mail address for first
%    author}} \and
%        Second author\inst{2}\fnsep\thanks{\email{Mail address for second
%             author if necessary}} \and
%        Third author\inst{3}\fnsep\thanks{\email{Mail address for last
%             author if necessary}}
%        % etc.
}

\institute{CSSM and ARC Centre of Excellence for Particle Physics at the 
Terascale, School of Chemistry and Physics, University of Adelaide SA 5005,
Australia} 
%\and
%           the second here 
%\and
%           Last address
%          }

\abstract{%
We present a vision for the next decade of hadron physics in which 
the central question being addressed is how one might win new physical 
insight into the way hadronic systems work. The topics addressed include 
the relevance of model building, the role of spontaneously broken 
chiral symmetry, spectroscopy, form factors and physics in the 
deep inelastic regime.
}
\maketitle
\section{Introduction}
\label{intro}
At times physics can be a very hard way to make a living. There can 
be few professions where one can work hard all day and go home feeling  
more stupid than when the day began. And yet, there are times when it is 
all worthwhile, when it really was worth the effort of getting out of bed. 
What makes it worthwhile, instills in us a sense of real achievement, 
is the feeling that one has actually won some new insight into how 
Nature works. My vision for the next decade is that, as a community, we 
will develop a clearer and more satisfying picture of the structure 
of hadrons and nuclei within the framework of QCD. This quest will, 
of course, involve new data and calculations of ever higher precision 
but, more than that, it will involve new physical insight and understanding.

\section{Models}
\label{sec-models}
There are some who are less than thrilled by the use of models in hadron 
physics and yet they are of fundamental importance. Flawed as all models 
necessarily must be, they are a vital part of the machinery needed to 
imagine new ideas and new ways of investigating Nature. Rather than arguing 
in generalities I briefly recall some examples of insights arising from 
models that have proven critical in motivating new experiments and 
guiding the development of our current understanding of hadron 
structure.

\subsection{Flavor asymmetries}
Although Sullivan and Feynman~\cite{Sullivan:1971kd} 
had suggested in the early 70's 
that the pion cloud of the nucleon 
might be significant in deep inelastic scattering (DIS), 
it would be fair to say 
that this was not taken at all seriously a decade later. Then, motivated 
by the cloudy bag model (CBM)~\cite{Theberge:1980ye,Thomas:1982kv}, 
their approach was used to explain  
the known asymmetry between strange and non-strange sea quarks in the 
nucleon~\cite{Thomas:1983fh}, most particle physicists ignored the work. 
Everyone knew that 
partons were non-interacting on the light-cone and this was generally taken 
to mean that clusters, such as a highly correlated $q-\bar{q}$ pair 
in a virtual pion, could not be relevant to deep inelastic scattering (DIS). 
Yet that same calculation 
also showed that the dominance of the $\pi^+ - n$ component in the chiral 
structure of the proton implied an excess of $\bar{d}$ over $\bar{u}$ 
quarks in the proton sea. Again this was ignored.

A decade later it was discovered that there was a very large violation 
of the Gottfried sum rule~\cite{Amaudruz:1991at} and later 
Drell-Yan experiments~\cite{Baldit:1994jk,Hawker:1998ty}, followed 
by semi-inclusive DIS 
measurements~\cite{Ackerstaff:1998sr}, 
confirmed that this originated in an excess of $\bar{d}$ over 
$\bar{u}$ quarks consistent with the 1983 
prediction~\cite{Melnitchouk:1991ui,Henley:1990kw,Melnitchouk:1998rv}.
The pion cloud of the nucleon, which had led to the 
prediction of precisely such an effect slowly began to be taken 
more seriously and new experiments are still being planned to explore 
the phenomenon in detail~\cite{Peng:2014hta}. 

Investigations of the structure functions of the nucleon within models 
such as the MIT bag~\cite{Schreiber:1991tc} 
and the chiral quark soliton model~\cite{Diakonov:1998ze} predicted 
non-trivial polarization of the anti-quarks in the proton sea arising from 
the modification of the vacuum inside a hadron, with $\Delta \bar{u} >0$ 
and $\Delta \bar{d} <0$. Experiments underway at RHIC have the capacity to 
tell us whether indeed this is the way Nature works. 

There is an almost universal assumption of charge 
symmetry~\cite{Londergan:2009kj,Londergan:1998ai} 
(e.g. $u \equiv u^p = d^n$ ) in the 
literature concerned with parton distribution functions. Indeed, 
phenomenological fits allowing for the possibility of charge symmetry 
violation (CSV) were only initiated a decade ago~\cite{Martin:2003sk}. 
Yet bag model 
investigations some 20 years ago had unambiguously predicted the sign 
and magnitude of CSV in the valence quark 
distributions~\cite{Sather:1991je,Rodionov}. The neglect 
of those predictions which, at least for low moments, have been 
confirmed by lattice QCD calculations~\cite{Cloet:2012db} in just the last few years
led to an over-inflated view of the importance of the deviation from 
the Standard Model expectations in neutrino-nucleus DIS. 

Later studies of the NuTeV anomaly have also revealed, again within 
a QCD motivated model of the EMC effect, the possibility of an 
additional source of difference between $u$ and $d$ valence quarks in 
nuclei with $N \neq Z$. This difference, called the iso-vector
EMC effect~\cite{Cloet:2009qs}, is a consequence of 
the iso-vector nuclear force acting 
on the bound quarks. Again, this insight from a model has suggested a number 
of experiments, including parity violating 
DIS on nuclei~\cite{Cloet:2012td}, which 
have the capacity to establish the phenomenon.  

\subsection{Spin}
The proton ``spin crisis'' resulting from the 
EMC measurements~\cite{Ashman:1987hv,Aidala:2012mv} of a large  
violation of the Ellis-Jaffe sum rule have inspired theorists to find
ways to explain it. Within a few months of the experimental paper two very 
different approaches were proposed. The theoretical beauty of the axial 
anomaly in QCD led to a rush to explore the possibility that at a scale of 
a few GeV$^2$ gluons might carry as much as 4 units of angular momentum, 
which through the box diagram containing the axial anomaly would resolve 
the crisis~\cite{Altarelli:1988nr}. 
No-one ever suggested how such an enormous gluon spin fraction 
might arise and eventually a new generation of polarized pp collider 
measurements have shown that indeed it does not exist~\cite{Adare:2008aa}.

The alternative suggestions based on rather well understood hadronic 
physics, which were initially left in the dust now appear to be correct. In 
particular, when a pion is emitted the proton tends to flip its 
spin, $p \uparrow \rightarrow \pi^+ + n \downarrow$, and within the 
CBM it was shown that this naturally accounts for half of the 
modern discrepancy~\cite{Schreiber88}. 
In addition, within essentially all QCD inspired 
quark models, the exchange of a single gluon is an essential part of the 
machinery needed to explain the hadron spectrum. For example, within the 
MIT bag model the $N-\Delta$ and the $\Sigma-\Lambda$ splitting
both arise from the one-gluon-exchange 
hyperfine interaction~\cite{Chodos:1974pn}. This 
same interaction {\em required by spectroscopy} naturally reduces 
the fraction of spin carried by the quarks~\cite{Myhrer}. 
Combining these two 
effects fully accounts for the current experimental 
value~\cite{Myhrer:2007cf} and lattice results~\cite{Thomas:2008ga}.

Of course, the ``experimental value'' is not independent of theory, 
because one must subtract the octet component of the nucleon spin ($g_A^8$)
from the experimental data to obtain the singlet spin fraction. Again, 
recent studies within the CBM suggest that SU(3) symmetry is broken 
by as much as 20 \% in this case~\cite{Bass:2009ed} 
and that has a significant effect 
on the deduced value of the proton spin fraction. Incidentally, since 
many fits to spin dependent PDFs impose an assumed value of $g_A^8$, 
this also has implications for the phenomenological spin dependent 
PDFs.

\subsection{Lattice QCD}
As lattice QCD has developed into a reliable, quantitative tool for 
calculating hadron properties within QCD one might imagine that models 
would be irrelevant. Yet that is far from correct. In the quest to 
understand how QCD works, lattice studies have created a new dimension 
to that occuring in Nature, namely the study of hadron properties as a 
function of quark mass. Initially, even for ground state masses,
it was essential to work with unphysical quark masses and to extrapolate 
to the physical quark masses in some way. This is still true 
now for many form factors and excited state properties. 
The CBM provided vital 
guidance in early studies of this kind~\cite{Leinweber:1999ig} 
and eventually inspired what 
has become known as finite range regularization (FRR)~\cite{Young:2002cj,Young:2002ib}. 
This approach has provided a natural explanation of the general absence 
of chiral curvature at quark masses above 40-50 MeV in {\em all hadronic 
observables}. It led to the precise calculation of the electric and magnetic 
strange form factors of the nucleon~\cite{Leinweber:2006ug,Leinweber:2004tc} 
and provided a ``back of the envelope'' 
understanding~\cite{Thomas:2005qb} of why those are so remarkably small.

\section{Spectroscopy}
\label{sec-spect}
As we hinted above, the situation with respect to excited states calculated 
using lattice QCD is far more fluid. The JLab and CSSM groups have made 
enormous progress at light quark masses two or three times the physical 
value, with as many as 4 or more excited states found for a given set of 
quantum numbers~\cite{Edwards:2012fx,Mahbub:2012ri}. 
It is still far too early to draw conclusions concerning 
the pattern of these excitations with respect to experiment. Nevertheless, 
this area will move rapidly over the next few years. As an indication 
of just what may be possible in terms of insight we mention the study 
of the wavefunction of the Roper resonance 
by Leinweber and collaborators~\cite{Roberts:2013ipa}, 
which clearly shows the node indicative of a $2s$ excitation. Since 
the Roper, along with the $\Lambda(1405)$, has been a bugbear in 
spectroscopy for decades (e.g. as to whether it is a multi-quark state, 
involves gluonic excitation, etc.), direct pictures of the distribution 
of quarks within the state should prove extremely valuable in 
solving the mystery.

The second major issue in hadron spectroscopy that must be mentioned 
is the existence, or otherwise, of exotic hadrons. The JLab lattice 
studies strongly suggest that these exist at an excitation energy perhaps 
0.8 to 1.0 GeV above the corresponding non-exotic 
hadrons~\cite{Edwards:2011jj}. This is 
consistent with the cost in terms of energy found in the MIT bag or 
much more recently within the Dyson-Schwinger formalism~\cite{Qin:2011dd}. Clearly the 
major advances in this area must initially be tied to the success of 
the experimental searches, for example at JLab 12 GeV.

\section{Form Factors}
No discussion of this topic at the present time would be complete without 
a mention of the mystery surrounding the charge radius of the proton 
following the muonic-hydrogen measurement of Pohl and collaborators~\cite{Pohl:2010zza}. 
At the present time the discrepancy between that beautiful measurement 
and the CODATA value is not understood at all and there is a distinct 
possibility that further investigation may reveal a hint of new physics, 
perhaps related to the muonic $g-2$ discrepancy. In this case there is 
no model we can point to for inspration.

At higher $Q^2$ the JLab discovery of an anticipated decrease in the ratio 
of the electric to magnetic form factors of the proton is of great 
interest~\cite{Perdrisat:2006hj}. 
We look forward to seeing data at even higher $Q^2$ to show 
{}for certain whether or not $G_E$ passes through zero. Lattice QCD 
has not yet reached these exulted values of momentum transfer but 
there has been some notable progress 
below 1.5 GeV$^2$~\cite{Shanahan:2014uka}. In terms 
of a deeper understanding of the physics in the region 5--10 GeV$^2$, 
the recent link between the behaviour of $G_E/G_M$ and the transition 
between non-perturbative and perturbative behaviour in the quark 
propagator suggested in a recent Dyson-Schwinger equation calculation 
makes it very interesting indeed~\cite{Cloet:2013gva}.

We already mentioned the success in both measuring and calculating 
the strange electric and magnetic form factors of the proton, which 
have occupied many people for the last two decades. This is a vital test 
of our understanding of non-perturbative QCD, in close analogy with 
the standing of the Lamb shift in QED. This because these form factors 
have their origins in so-called ``disconnected diagrams'', or 
non-valence physics, just like vacuum polarization. Pushing these 
tests to higher precision is at a stand still for the time being 
as one looks for new experimental techniques and tries to pin 
down the extent of CSV in the nucleon elastic form factors.

\section{The Deep Inelastic Regime}
\label{sec:DIS}
Over the past 40 or more years unpolarized DIS measurements have 
managed to define very precisely many important properties of the 
PDFs of the nucleons. There are cases where physics demands at the 
LHC may need more but, at least for unpolarized scattering, it is 
in the area of flavor asymmetries that the most interesting 
questions remain open. We have already mentioned most of these 
issues in sect.~2. Semi-inclusive DIS is one promising technique to 
explore such questions but, especially once it comes to strange 
quarks and fragmentations functions that involve koans or unfavoured 
fragmentation there is much we do not know. Model studies may well 
provide useful information there~\cite{Matevosyan:2010hh}.

However, it is in the measurement of processes involving polarization 
that there is the greatest activity. The next decade will see a revolution 
in our understanding of the Collins and Sivers effects, of deeply 
virtual Compton scattering and so on. Perhaps because we are at such 
an early stage in this work, there is a great deal of effort needed 
to build better models that more accurately reflect the consequences 
of QCD itself. Eventually we may hope for a much deeper  
understanding  of the role of orbital angular momentum in hadrons.
We already know that the conversion of valence quark spin into quark 
orbital angular momentum, carried by pions and anti-quarks, is the 
explanation for the EMC spin crisis, but the finer details demand 
much more effort.

\section{Conclusion}
\label{sec:end}
In this very limited space it has been impossible to do more than set 
out a rough sketch of a vision for hadronic physics over the next 
decade. New facilities such as JLab at 12 GeV, GSI-FAIR, J-PARC 
and one or more electron-ion colliders, backed by upgrades at RHIC, 
will provide essential new data. Lattice QCD will continue to grow in 
reliability and accuracy. We will probe the QCD structure and origins 
of atomic nuclei in new ways. Yet, in the end, our success will be judged 
by the new insights into how hadrons work, the new paradigms that are 
established. It will be the leaps in our qualitative understanding, many 
related to the beauty of new models that capture the essence of the 
hadronic systems that will stand out.
These are exciting times.

\section*{Acknowledgements}
This work was supported by the University of Adelaide and the Australian 
Research Council through an Australian Laureate Fellowship and through 
the ARC Centre of Excellence in Particle Physics at the Terascale.

%XXXXXXXXXXXXXXXXX
%

\end{document}